\documentclass[useAMS,usenatbib,aas_macros]{mn2e}
\usepackage{aas_macros}
\usepackage{amsmath}
\usepackage{graphicx}% Include figure files
\usepackage{xspace}
\usepackage{amssymb}

\newcommand{\equ}[1]{eq.~(\ref{eq:#1})}

\newcommand{\Equ}[1]{Eq.~(\ref{eq:#1})}

\newcommand{\se}[1]{\S\ref{sec:#1}}
\newcommand{\fig}[1]{Fig.~\ref{fig:#1}}

\newcommand{\Fig}[1]{Figure~\ref{fig:#1}}

\newcommand{\be}{\begin{equation}}
\newcommand{\ee}{\end{equation}}
\newcommand{\bea}{\begin{eqnarray}}
\newcommand{\eea}{\end{eqnarray}}

\newcommand{\mnc}{C\xspace}
\newcommand{\mcl}{CH\xspace}
\newcommand{\mmc}{CHC\xspace}

\newcommand{\msun}{M_\odot}

\newcommand{\ifm}[1]{\relax\ifmmode#1\else$\mathsurround=0pt #1$\fi}
\newcommand{\kms}{\ifmmode\,{\rm km}\,{\rm s}^{-1}\else km$\,$s$^{-1}$\fi}
\newcommand{\ergsec}{\,{{\rm erg\,sec}^{-1}}}

\newcommand{\kpc}{\,{\rm kpc}}
\newcommand{\pc}{\,{\rm pc}}
\newcommand{\Gyr}{\,{\rm Gyr}}
\newcommand{\yr}{\,{\rm yr}}
\newcommand{\keV}{\,{\rm keV}}
\newcommand{\ltsima}{$\; \buildrel < \over \sim \;$}
\newcommand{\lsim}{\lower.5ex\hbox{\ltsima}}
\newcommand{\gtsima}{$\; \buildrel > \over \sim \;$}
\newcommand{\gsim}{\lower.5ex\hbox{\gtsima}}
\newcommand{\prop}{\propto}

\def\sy{\,M_\odot\, {\rm yr}^{-1}}

\usepackage{color}

\title[Gravitational Heating in Clusters by Clumps]
{Gravitational Quenching by Clumpy Accretion in Cool Core Clusters:
Convective Dynamical Response to Overheating}

\author[Y. Birnboim, A. Dekel]
{Yuval Birnboim$^{1,2}$, Avishai Dekel$^{2}$, \\
\\
$^{1}$Harvard Smithsonian Center for Astrophysics, 60 Garden Street,
Cambridge, MA 02138 USA\\
$^2$Racah Institute of Physics, The Hebrew University, Jerusalem 91904 Israel\\
ybirnboim@cfa.harvard.edu; dekel@phys.huji.ac.il}

\begin{document}

\large

\pagerange{\pageref{firstpage}--\pageref{lastpage}} \pubyear{2010}

\maketitle

\label{firstpage}

\begin{abstract}
Many galaxy clusters pose a ``cooling-flow problem", where the observed X-ray emission from  
their cores is not accompanied by enough cold gas or star formation.
A continuous energy source is required to balance the cooling rate over 
the whole core volume. We address the feasibility of a gravitational heating
mechanism, utilizing the gravitational energy released by the gas that streams 
into the potential well of the cluster dark-matter halo. We focus here on a 
specific form of gravitational heating in which the energy is transferred to 
the medium thorough the drag exerted on inflowing gas clumps.
Using spheri-symmetric hydro simulations with a subgrid representation
of these clumps, we confirm our earlier estimates that in haloes 
$\geq 10^{13}\msun$ the gravitational heating is more efficient than
the cooling everywhere. The worry was that this could overheat the core and
generate an instability that might push it away from 
equilibrium. 
However, we find that the overheating does not change the global halo 
properties, and that convection can stabilize the cluster by carrying energy 
away from the overheated core. 
In a typical rich cluster of $10^{14-15}\msun$, with $\sim 5\%$ of the accreted
baryons in gas clumps of $\sim 10^8\msun$, we derive upper and
lower limits for the temperature and entropy profiles and show that
they are consistent with those observed in cool-core clusters. We
predict the density and mass of cold gas and the level of 
turbulence driven by the clump accretion.
We conclude that gravitational heating is a feasible mechanism for
preventing cooling flows in clusters.
\end{abstract}


%%%%%%%%%%%%%%%%%%%%
\section{Introduction}
\label{sec:intro}

%cooling flow problems
Galaxy clusters can be divided into two distinct populations according to
the X-ray luminosity of their central cores
\citep{sanderson06}. Cool-core (CC) clusters are centrally 
concentrated, highly luminous in X-ray, and have central cooling times
of $0.1-1\Gyr$. Non-cool-core (NCC) clusters have lower
densities and luminosities near the centre, and their central cooling
times are typically a few Gyrs  \citep{donahue06}. 
The population of CCs has little
internal variability, and they all exhibit a typical temperature
profile with a decline by a factor of 2-3 in the innermost few 
$10\kpc$ \citep{leccardi08}. 
Based on the short cooling time in CCs, one expects to observe 
gas in intermittent temperatures, a high star-formation
rate in the brightest central galaxy (BCG) ($\gtrsim 100\msun/\yr$), 
and a large stellar mass in the BCG ($10^{12-13}\msun$),
none of which is observed.  
These are three manifestations of the cooling-flow problem in clusters 
\citep{fabian94}.  
Since the cooling time
is inferred directly from observations based on the luminosity and
temperature, the discrepancy between the
expected cooling rates and the gas that actually cools indicates
that some heating mechanism is keeping the gas hot in a stable 
configuration.

% AGN feedback
Several mechanisms have been proposed as potential solutions to
the cooling-flow problem. Most popular is AGN feedback, where energy
or momentum are provided by an active galactic nucleus
either in an intense quasar mode \citep{ciotti07}, in a slower 
radio mode \citep[][for a review]{best07,cattaneo09},
or via cosmic rays \citep{guo08}. 
The AGNs clearly release sufficient power to balance the cooling in the cores,
and the observed big radio and X-ray bubbles in some cluster cores
\citep{birzan04} is likely evidence for AGN feedback. However,
the coupling of the AGN energy to the gas in the whole core 
volume is not easily understood,
and the requirement of continuous heating is not a trivial constraint
\citep{deyoung08}. 
Another possibility is that
the cooling instability might be locally suppressed with the addition of
non-thermal pressure sources such as cosmic rays or 
turbulent magnetic fields \citep{sharma10}.

% gravitational heating
Here we utilize the fact that the gravitational power
released as fresh baryons stream into the potential well created by a 
massive halo is more than enough to balance the cooling rate at the centre 
\citep{fabian03,db08,wang08}.
Among the mechanisms through which this energy is transferred to the inner 
halo one could consider dynamical friction 
\citep{elzant04b,faltenbacher07b,naab07,khochfar08,johansson09}, 
thermal conduction \citet{kim03}, 
or turbulence \citep{narayan04}.
In particular, the turbulence induced by accreting sub-structures may 
disturb the magnetic field in a way that could make the conduction more 
efficient \citep{parish10,ruszkowski10}. 
 
%clumps
The gravitational power that is released during the streaming of baryons
into clusters of $M \geq 10^{13}\msun$ is indeed sufficient for
balancing the expected radiative losses \citep[][hereafter
DB08]{db08}\footnote{We note that the overcooling problem is less
  pronounced for haloes between $10^{12}-10^{13}\msun$ \citep{bdn07}.}.
The challenge is to deposit this energy at the inner cluster core of 
$\sim 30-100\kpc$, where most of the cooling takes place. 
The energy should be evenly distributed over the whole core volume
and continuously over several Gigayears.
It should be performed in a way that is consistent with 
the observed entropy profile \citep{donahue06} and metallicity
\citep{rebusco06} profile. Here we
investigate a model in which some of the accreted gas is in dense
and cold ($\sim 10^4$K) gas clumps.
These clumps do not stop at the virial shock but rather penetrate to the 
inner parts of the halo. The continuous accretion of many clumps 
through the halo can distribute the energy smoothly in a large volume,
as required, unlike AGN feedback that is episodic and originates from
a very small region near the black hole. 
The main physical mechanism that couples 
the clumps and the halo gas is hydrodynamic drag, subsonic or supersonic,
which decelerates the clumps and causes the deposition of
their kinetic and potential energy in the ambient medium.
Hydrodynamic drag is more efficient when the clumps are smaller and moving
faster, as opposed to dynamical friction that becomes more efficient for 
more massive clumps and for transonic velocities \citep{ostriker99}.
We pointed out in DB08 that
the gravitational heating is likely to be associated with the streams 
that build the cluster along the filaments of the cosmic web
\citep{db06,dekel09}. These streams
could transfer their energy to the cluster core via the generation of
turbulence and other mechanisms that are not necessarily associated with 
cold clumps \citep{zinger11}. Still, the heating through cold clumps is
a concrete example that allows a simple study of the dynamical response 
with general implications that are not limited to this particular coupling
mechanism.

%summary of db08
In DB08 we showed that accreted clumps could balance the cooling
in haloes more massive than $6\times 10^{12}\msun$,
provided that the clumps contain a non-negligible fraction of the infalling
gas ($\geq 10\%$ for a $10^{15}\msun$ halo), and that the clump masses are
in the range $10^4 - 10^8\msun$. The clumps heat the intra-cluster medium (ICM)
via drag until they disintegrate by hydrodynamical 
instabilities \citep{murray04,maller04}. 
The permitted mass range for the clumps took into account additional
criteria for clump survivability,  
such as Bonnor-Ebert instability, conduction and
evaporation.
When the clumps are sufficiently massive, they survive long enough to
penetrate through the outer halo and reach the 
centre, while by being not too massive, their interaction with the
ICM is sufficiently strong for most of the
clump energy to be deposited in the core within a Hubble time.
The basic features studied in DB08 assuming a static system
will be implemented below in a dynamical evolving configuration.

%overheating instability 
The presnt study is motivated by a potential {\bf overheating problem.}
The heating rate by drag (and by other mechanisms such as dynamical
friction, cosmic rays from AGN)
is proportional to the density of the hot ambient gas,
$\dot{e}_{\rm heat} \propto \rho_{\rm hot}$. 
On the other hand, the Bremsstrahlung cooling rate 
per unit volume scales like $\dot{e}_{\rm cool} \propto \rho_{\rm hot}^{1.5}$ 
(assuming isobaric gas). 
This generates an {\bf instability} through a positive feedback loop 
\citep{field65,conroy08}.
A small negative density perturbation, 
e.g., produced by gas expansion due to overheating, 
would make the ratio of heating to cooling rate increase as
$\dot{e}_{\rm cool}/\dot{e}_{\rm heat}\propto \rho_{\rm hot}^{0.5}$,
leading to more overheating, enhanced pressure, and a runaway expansion.
An analysis of the consequences of this instability, 
and an investigation of possible mechanisms that could keep the cluster
in equilibrium, necessitate a dynamical treatment.

%necessity of convection 
This unstable overheating, as reproduced in spheri-symmetric simulations below, 
results in expanding shells that heat to temperatures as high as $10^9K$,
clearly at odds with observations. In the real world, this overheating
must be regulated by processes that smooth 
temperature or entropy gradients by heat transfer 
through conduction or convection. 
Conduction is suppressed in the presence of magnetic fields 
\citep[][and reference within]{fabian94}, though it might be boosted
almost to its 
maximum possible value 
\citep{spitzer62} by turbulence \citep{narayan01,balbus08,parish10,ruszkowski10}, 
which might be a natural product of clump heating.
{\bf Convection} is a promising mechanism for smoothing the
local instabilities. In the simple case of an ideal gas with a uniform chemical
composition in a spherical potential well, convection
occurs in regions where the entropy is declining with radius. 
However, the strength of this convection is uncertain, as it depends on the 
gradients in gas properties and on the magnetic fields.
Weak magnetic fields make the gas more susceptible to convection, 
with the entropy-gradient criterion replaced by the temperature gradient
\citep{balbus00,balbus01,quataert08}, but for certain types of perturbations, 
the convection strength might be drastically suppressed
by effects related to magnetic tension \citep{parrish08b,parrish09}.
Regardless of the actual energy transport mechanism, one expects 
nature not to permit shells of $\sim 10^9$K in close
contact with shells of $\sim 10^7$K, and to act to smooth such a discontinuity. 
Motivated by the turbulence that is expected to be generated by the clumps,
we focus below on convection as the mechanism that smooths steep
gradients.
%XXX,
%but we expect the final profiles to be qualitatively similar when the mechanism for
%energy transport is conduction because smooth entropy and pressure
%(the result of convection) is equivalent to smooth temperature and pressure 
%(the result of conduction).

% mixing length convection
In this paper, we mimic this smoothing process by a 1D mixing-length 
convection model \citep{spiegel63}, with
a the mixing-length coefficient $L$ the single free parameter. We find
that the results are almost independent of the value of this parameter 
as long as noticeable convection occurs. This allows us to further simplify
the model by assuming that the convection is maximal, namely 
the energy transfer rate is limited by the requirement that hot
bubbles accelerate until they become supersonic, at which point 
they disintegrate. This leaves us with no free parameters in our 
convection model. 

%dynamic evolution
There are some additional benefits from a dynamic treatment of the
clump heating process. The analysis of DB08 considered simple Monte-Carlo 
clump trajectories within an otherwise static halo in hydrostatic equilibrium.
This assumption of a static halo could be valid for one Gigayear but the
cluster may evolve considerably over a Hubble time,  due, for example, to the gradual
increase of virial temperature and the growth of the BCG.  
Furthermore, if clump heating is taking place, cold
clumps continually get destroyed near the halo centre, dumping
cold gas near the BCG. This dilution of the hot gas with cold gas is not 
a problem for the heating-cooling balance because the clumps bring in 
several times the energy needed for heating themselves to the cluster virial 
temperature, but a proper account of this clump deposition requires
a dynamical analysis of an evolving cluster.

%structure of paper
In \se{numerics} we describe the implementation of the clump model and
of the convection model in the 1D hydrodynamic code. 
In \se{results}
we show results of hydrodynamic cluster simulations with convection
and clump heating that match the observed temperature and entropy profiles of
clusters and the cooling rates in clusters without a need for any
additional feedback. 
In \se{ncl} we address possible direct and indirect observations of the cold
clumps.
In \se{discussion} we summarize and conclude.

%%%%%%%%%%%%%%%%%%%%%
\section{Methods}
\label{sec:numerics}

%------------------------
\subsection{Implementing Clumps in 1D Hydrodynamic Simulations}
\label{sec:clump_numerics}

%clump physics
According to our estimates in DB08, heating by clumps requires clump masses 
in the range $10^4-10^8\msun$. In order to properly resolve drag
forces and clump disintegration via hydrodynamical instabilities,
these clumps should be resolved by at least 1000 cells or SPH
particles (i.e., 10 cells across each dimension). The implied required
dynamical range in a cluster of $10^{14}-10^{15}\msun$ is impractical,
so simulation of such clumps requires a sub-grid model.
We develop such a model below, and describe its
implementation in a 1D spherical hydrodynamical code.

The clumps are made of cold and partly ionized gas at $\sim 10^4$K 
in pressure equilibrium with their surrounding hot halo. 
For a rich cluster of galaxies, with a virial temperature $\sim 10^7$K,
the overdensity within the clumps is about $10^3$.
The clumps couple to the hot gas by a drag force 
\be
f_{drag}=\frac{1}{2}\, C_{\rm d}\, A\, \rho_{\rm hot}\, v_{\rm rel}^2 \, ,
\label{eq:drag}
\ee
acting opposite to the relative direction of motion.
Here $C_{\rm d}$ is the drag coefficient ($\sim 1$ for a spherical gas clump), 
$A$ is the cross-section surface area of the clump ($\pi R_{\rm cl}^2$), 
$\rho_{hot}$ is the density of the hot component,
and
$v_{\rm rel}$ is the relative velocity between the clump and the halo gas.
\Equ{drag} holds for subsonic and supersonic motions, though the value of
$C_{\rm d}$ may vary, especially in the trans-sonic regime where it could
be a few.
The deceleration ($f_{drag}/m_{cl}$) is proportional to the ratio of clump
surface area to volume, $\prop m_{cl}^{-1/3}$. 
This dictates lower and upper limits to the relevant clump masses.
Clumps that are too small cannot penetrate through the outer halo into the
core, and clumps that are to large cannot deposit a significant fraction
of their energy in the inner halo on a time scale shorter than the Hubble time.

Single clump simulations \citep{murray04} indicate that most of the
energy dissipated in this process goes into the hot ambient gas.   
The survivability of these clumps is an open question as they could
be destroyed by many different effects.
This has been discussed in DB08 and in \citet{maller04}. 
In a nutshell, if the clump is more massive than
$\sim 10^8\msun$ it would exceed the Bonnor-Ebert critical mass 
(the equivalent of the Jeans mass for pressure-confined spheres) 
and it would collapse under its self-gravity and turn into stars.
If the clump is less massive than $\sim 10^4\msun$, evaporation and conduction 
are expected to disintegrate the clump. 
Hydrodynamic instabilities, particularly Kelvin-Helmholtz
instability, tend to break the clump, typically after it has repelled
the equivalent of its own mass in ambient gas \citep{murray04}. 
The clump masses then cascade down toward the lower limit for clump mass.

%-------------------------
\subsubsection{The Hydra Code}

Subgrid recipes of the effects mentioned above were incorporated into the 1D
spherical code Hydra \citep{bd03}. The code is finite-difference
Lagrangian with von-Neumann first and second order artificial
viscosity. Dark matter is described as zero-width shells that
propagate through the gaseous shells, and interact with them
gravitationally. The coupled density fields of gas and dark matter shells
are propagated using a 4th order Runge-Kutta method. Time steps are
defined by the minimum of the Courant conditions and the allowed
deviation of the forth order scheme from fully explicit (1st order)
time step. When this difference exceeds some preset epsilon, the
previous values are restored, and a new step with decreased time step
is performed. A comprehensive description of the hydrodynamic and
dynamic equations for the gas and the dark matter shells, 
the numerical scheme used and convergence tests and a comparison to
analytic test problems can be found in section 3 and the appendixes of
\citet{bd03}.

The baryons and the dark matter shells are assigned
a preset angular momentum that is added as an impulse when the shell
is at its turn-around. This prevents a numerical and physical
singularity at the centre. The angular momentum of the dark matter is
set so that the rotational kinetic energy
is $18\%$ of the radial kinetic energy at the virial radius.
The results are insensitive to
this choice. The baryonic angular momentum is set to
produce a spherical, angular-momentum supported ``disc'' of radius
$\sim 10\kpc$ for a cluster halo of $10^{15}\msun$, to mimic a BCG. 
Since centrifugal forces scale like $r^{-3}$, the angular momentum of 
the baryons is negligible at a distance comparable to a few disc
radii above the disc. We note that a spherical code is not the right
tool for studying disc formation, and this setup is essentially an inner
boundary condition for the halo simulation. 
In addition, the code imposes a central smoothing length on the gravitational
acceleration, $a_{\rm g}=G\,M/(r+s)^2$, typically with $s \sim 50\pc$.
Radiative cooling is calculated by a metallicity dependent
cooling function \citep{sutherland93} with a constant preset metallicity.

The initial conditions, in terms of shell masses, radii and peculiar 
velocities,
were set at $z=100$ such that the future accretion rate onto the growing halo
will follow a desired accretion rate \citep[][appendix C]{dekel81,bd03}.
Specifically,
the initial perturbation used here yields an accretion history that
traces that of an average main progenitor according to the EPS 
approximation \citet{press74,lacey93,neistein06}. 
The procedure is described in detail in
\citet{bdn07}. The code has been compared successfully to analytic 
predictions of \citet{bertschinger85} and to a Von-Neumann-Sedov-Taylor 
problem.
The Courant conditions and epsilons are set
so that the global energy conservation over a Hubble time is always
better than $1\%$ and the spatial convergence was tested for each set of
simulations. Timesteps are consequently $\sim 10^{-5}\Gyr$
throughout most of the simulation.

%------------------------
\subsubsection{Drag forces in 1D}
\label{sec:drag}
The subgrid model for clumps is similar in its approach to ``sticky
particle'' techniques in the sense that it calculates subgrid
interactions on otherwise N-body particles. Here we
define ``clump-shells'', which, like dark matter shells, are able to
penetrate through baryonic shells. The clump shells are assigned some angular
momentum (similarly to the baryon and dark matter shells) that stops
them from reaching the singularity at the 
centre. Clump shells typically oscillate around the halo's centre
before the processes described below destroy them.

Each shell is assumed to contain
$n_{\rm cl}$ clumps with mass 
\be
m_c=\frac{M_{\rm shell}}{n_{\rm cl}}.
\label{eq:mc}
\ee
 $M_{\rm shell}$ and $m_c$ are the total shell mass and the mass of each
 gaseous clump respectively. The shells interact with the
baryons by decelerating according to \equ{drag}. The drag force equation
and energy equation of an interaction between some clump $i$ and a parcel of
gas are:
\be
f^i_{\rm cl}=-F^i_{\rm gas},
\ee
and
\be
\dot{E^i}=f^i_{\rm cl}\,v^i_{\rm cl}+F^i_{\rm gas}\,u_{\rm gas}+Q^i=0,
\ee
respectively with $f^i_{\rm cl}$ and $F^{i}_{\rm gas}$ the forces on the
clump and gas parcels arising from the clump-gas interactions,
and $v^i_{\rm cl}$ and $u_{\rm gas}$ the velocities of clump $i$ and the
gas respectively. $Q^i$ is the rate at which clump $i$ heats the gas
parcel it is embedded in at that time.
We assume that many clumps are present within each parcel of gas, and
that their motions are isotropic so their forces cancel out:
\be
\sum F^i_{\rm gas}=0\, ,
\ee
and

\begin{align}
\dot{E}_{\rm tot}&=\sum f^i_{\rm cl}\,v^i_{\rm cl}+\sum F^i_{\rm gas}\,u_{\rm gas}+\sum Q^i
\label{eq:total_e}\\
&=\sum f^i_{\rm cl}\,v^i_{\rm cl}+\sum Q^i=0\, .\notag
\end{align}

In the reference frame of a gas parcel , the clump loses energy, which is
converted to heat, so using \equ{drag} we identify 
\be
Q^i=f_{\rm drag}\,v_{\rm rel}=\frac{1}{2}\,C_d\,A\,\rho_{\rm
  hot}\,|v_{\rm rel}|^3\, ,
\label{eq:q}
\ee
as the heating rate that clump $i$ heats the gas parcel in which it is
embedded.
The total energy of the clumps and of the gas should be conserved on
average according to \equ{total_e}, assuming there are enough clumps
so the averaging is correct, and that the assumption that the clumps
have isotropic velocities is good. The difference equations in Hydra conserve energy
algebraically\footnote{an algebraic scheme is such that the exact
  energy term is always added to one component and subtracted from the
  other explicitly, making sure the energy is conserved to the machine
  accuracy}(making energy conservation independent of resolution), and we do
not want to violate this property. We thus require a
detailed balance between the clump deceleration and energy deposition
of each clump:
\be
f^i_{\rm cl}\,v^i_{\rm cl}=Q^i\, ,
\ee 
which implies:
\begin{align}
f_{\rm cl}&=f_{\rm drag}\frac{v_{\rm rel}}{v_{\rm cl}}\notag\\
&=- \frac{1}{2}\,C_d\,A\,\rho_{\rm hot}\,v_{\rm
  rel}^2\frac{|v_{\rm rel}|}{v_{\rm cl}}\label{eq:fclump}\\
&=-\frac{1}{2}\,C_d\,A\,\rho_{\rm hot}\frac{|v_{\rm rel}|^3}{v_{\rm
    cl}}\, .\notag
\end{align}

\Equ{fclump} recovers the exact solution when the gas parcel is at
rest (which is the typical case of heating of a hydrostatic gas) and
when there is no relative velocity between the gas and the clump (i.e. no drag).

Physically, the drag forces always act to decrease the radial and tangential
components of the velocity of clumps. While the radial velocity is
replenished by the gravitational force, the tangential velocity
monotonically decreases in time, so clumps lose angular momentum,
becoming more radial in their 
trajectories. Since the angular momentum of the clump shells in our
simulations is conserved, the clumps in the simulations cannot spiral
towards the centre. To compensate for this problem, and allow clumps to
reach the central core, a much smaller angular momentum is assigned to
the clumps, placing them on almost radial trajectories.

\subsubsection{Fragmentation of Clumps}
Once the framework for accelerations and heating is defined, we
proceed to  implement recipes for clump evolution. In this work, we
implemented only the most crucial additional recipes from
DB08: clump fragmentation, and clump destruction. A clump
fragments into $n_{\rm frag}$ clumps (2 thorough out this work) once
 it has repelled its own mass of ambient gas. The amount of gas repelled is calculated by
numerically integrating over $\pi
r_{\rm cl}^2\,\rho_{\rm gas}v_{\rm rel}\,dt$ in the same Runge-Kutta scheme used for
the dark matter. After each timestep, the column
mass and clump mass is compared. When the column mass exceeds $m_{\rm cl},$
$m_{\rm cl}$ is divided by $n_{\rm frag}$, $n_{\rm cl}$ is multiplied by $n_{\rm frag}$,
and the column mass integral is reset to $0$. As $m_{\rm cl}$ becomes
exceedingly small, its drag deceleration 
become large, and the periods between consequent fragmentation events become
shorter. Clumps are destroyed when their mass
decreases below a critical mass, at which conduction and evaporation
is expected to disintegrate them completely \citep[in this work -
$10^4\msun$, ][]{db08}. 

\subsubsection{Destruction of Clumps}
\label{sec:destruction}
The destruction of clump is achieved by adding its mass, to the
corresponding baryonic shell. The velocity and angular momentum are
not changed, and the internal energy and temperature is calculated by
mixing the cold and hot components 
according to energy conservation by solving for the final internal energy,
$E_{\rm int}^f$ in:
\begin{align}
&M\,(E_{\rm int}+E_{\rm kin}+E_{\rm grav}) +m\,(e_{\rm int}+e_{\rm kin}+e_{\rm grav})=\notag\\
&(M+m)(E_{\rm kin}+E_{\rm grav})+(M+m)\,E_{\rm int}^f\, .
\end{align}
with $M,E$ correspond to the baryonic shell values, $m,e$ to the clump
shell. All energies are specific energies (per unit mass) and $e_{\rm int}=C_v\,T_{\rm cl}$ with $T_{\rm cl}=10^4K$.
Rarely,  the final temperature will drop below
$10^4K$, at which case the final temperature is set to $10^4K$. This
is found to occur only for baryonic shells that are cold (around
$2\times10^4K$) and on a free fall to the BCG, and the temperature floor
that is artificially applied
never stops their infall. In this case, the fictitious energy is
tracked throughout the run and is always negligible. The overall
accreted mass is not effected by this correction.

\subsubsection{Applicability to 3D Simulations}
\label{sec:3d}
In the 1D case described above, the gas parcel is a finite-dimension
gas shell, and the clump is a thin shell. While 3D simulations are
beyond the scope of this work, if one wishes to incorporate the
effects of cold clumps in 3D simulations, a generalization for the 3D
case is readily available, as follows. Both SPH and grid-based (Eulerian or
Lagrangian) simulations that model dark matter as N-body particles can 
assign clumps to a dark matter particle according to \equ{mc},
calculate its acceleration according to \equ{fclump} and heat the gas
according to \equ{q}. In cosmological simulations, it is also
necessary to self-consistently create those clumps. These can either
be created semi-analytically according to cooling instabilities
\citep{field65,binney09} or by identifying unresolved gaseous clouds
and replacing them with clump particles \citep[][identify
clump formation on Milky Way scales, but resolution would not allow to scale
this procedure to galaxy cluster scales]{keres09}. 

\subsection{Cell splitting - a 1D Adaptive Mesh Refinement}
\label{sec:amr}
In the Lagrangian formalism, the amount of baryonic mass within each
shell is constant throughout a simulation. Clump destruction,
however, violates that by dumping mass into the baryonic
shells at the event of clump destruction. This mass deposition is
expected to occur preferentially at specific regions. Indeed, the
simulations discussed below initially caused the formation of  a few
extremely massive shells that absorbed most of the mass of the clumps. This situation 
(of large contrasts between adjacent shells in mass and widths)
reduces the accuracy of the numerical scheme, and effects the
accretion rates onto the BCG. To overcome this, an
adaptive splitting of baryonic shells is performed: when the shell is
wider by some factor than both its neighbours, or when the shell's
width divided by its radius exceeds some preset fraction, the shell is split to
two. The factors
ultimately used were: $\Delta r_n>4 \times {\rm max}(\Delta r_{\rm n-1},\Delta
r_{\rm n+1})$ and $\Delta r_n/r_n>0.2$ (with $\Delta r_n$ and $r_n$ the
width, and central radius of shell $n$) which caused splitting to occur a
few dozens times throughout a full, Hubble time simulation.
This choice of parameters is a result of trial and error motivated by the requirement that
except for transient periods, the width of Lagrangian shells are
roughly equally spaced (locally), ensuring the robustness of the
Numerical scheme.  

A shell is split into two constant mass shells. Values that are
naturally defined at centres of shells (density, temperature) are
treated as step functions and their values are equal in the two new
shells. Values that are defined on boundaries (radius, velocity, angular
momentum) are interpolated linearly with mass. This definition ensures
that the total internal energy of the system remains the
same. The potential and kinetic energies, however, can change. This energy
non-conservation is not corrected for explicitly. Rather, it is
treated as non-conservation and tracked throughout the run. In the
high resolution simulations shown below (2,000 baryonic shells and
10,000 clump and dark matter shells), the total, overall energy is
conserved to better than $10^{-2}$ over a Hubble time.

\subsection{Convection and Mixing Length Theories}
\label{sec:mlt_numerics}
\subsubsection{Convection}
A long term balance between cooling and heating requires, in addition
to sufficiently large energy injection, a correct distribution of this
energy. The cooling rate, at constant pressure, scales as
$\dot{e}_{\rm cool} \sim \rho^{1.5}$, and most heating mechanisms (the drag forces
discussed here, dynamical friction, radiative heating from supernovae
and AGNs) generally heat according to $\dot{e}_{\rm heat}\sim \rho$. Even if at some
point in space, and at some initial time, $\dot{e}_{\rm cool}/\dot{e}_{\rm heat}=1$, this ratio
scales like $\rho^{0.5}$. This relation indicates a positive feedback
so if density increases, cooling is more efficient, causing a further
increase in density at constant pressure, and unstable cooling will
occur. Vice versa, a density decrease increases the relative
importance of heating over cooling, decreasing the density further,
and an over-heating instability occurs. This point has been made by
\citet{conroy08} \citep[see, however, ][who claim that residual
magnetic turbulence might inject energy into the gas in a stable manner]{kunz10} ,
using 1D hydrodynamic calculations of clusters initially hydrostatic within
a static potential well. They find that stable, long term equilibrium
requires fine tuning of the heating efficiency which is unlikely. The
present work differs in that it treats the gravitational drag feedback
and the cluster 
evolution from initial cosmological perturbation consistently, but
should still suffer from a heating instability. Such an overheating
will manifest as a 
shell or region of shells of gas continuously becoming hotter and
under-dense, with very 
high entropy. In a 3D configuration, this entropy inversion is unstable to
convection when entropy is declining outwards: a slightly under-dense
parcel of gas floats buoyantly, 
carrying energy and momentum outwards, and over-dense parts sink towards
the centre reducing the average entropy and specific energy of the
core. For our 1D simulation we invoke a 1D subgrid model for
convection - mixing length theory \citep{spiegel63}.
 
\subsubsection{Mixing Length Theory}
\label{sec:mlt}
Convection occurs when an adiabatic displacement of a parcel of gas,
in pressure equilibrium with its new position, results in a net force
on that parcel tending to increase the displacement. This requires
that, for an upward displacement, the temperature of the gas parcel
will be smaller than its surrounding:
\begin{align}
\Delta \nabla T&\equiv \left(\frac{\partial T}{\partial
    r}\right)-\left(\frac{\partial T}{\partial r}\right)_{\rm S}\notag\\
&=\left[\left(\frac{\partial T}{\partial S}\right)_{\rm P}\frac{\partial
    S}{\partial r}+\left(\frac{\partial T}{\partial P}\right)_{\rm S}\frac{\partial
    P}{\partial r}\right]-\left(\frac{\partial T}{\partial P}\right)_{\rm S}\frac{\partial
    P}{\partial r}\notag\\
&=\left(\frac{\partial T}{\partial S}\right)_{\rm P}\frac{\partial
    S}{\partial r}\, ,
\label{eq:ddt}
\end{align}
with the first term in the r.h.s corresponding to the actual
temperature derivative in the profile, and the second to the adiabatic change in
temperature as a result of the pressure profile of the halo. A
negative value of $\Delta \nabla T$ allows for convection to occur. It is
convenient to derive this relation using a form of the ideal equation
of state in which the two thermodynamic free parameters are the
entropy, $S$, and the pressure, $P$.
\begin{align}
P&=\frac{N_{\rm A}\,k_{\rm B}}{\mu}\,\rho\, T\, ,\notag\\
S&=\frac{N_{\rm A}\,k_{\rm B}}{\mu}\,\ln\frac{T^{3/2}}{\rho}\, ,
\end{align}
can be inverted  (setting $\hat{\mu}=\mu/N_{\rm A}\,k_{\rm B}$) into: 
\begin{align}
\rho&=\left(\hat{\mu}\,P\right)^{3/5}e^{-(2/5)\hat{\mu} S}\, ,\notag\\
T&=\left(\hat{\mu}\,P\right)^{2/5}e^{(2/5)\hat{\mu} S}\, .
\end{align}
Plugging these relation into the derivatives in \equ{ddt} we get:
\be
\Delta \nabla T=\frac{2}{5}\,\hat{\mu}\,T \frac{\partial S}{\partial
  r}\, ,
\ee
recovering the known results that when the composition of the gas is
constant, entropy inversion leads to convection.

The buoyant bubbles rise for a typical length
before being destroyed by Kelvin Helmholtz instability or
conduction. The details 
of this destruction depend on the size of the bubbles, the smoothness
of the density and gravitational profile, conduction and magnetic
fields. Solving
for it requires fine 3D simulation of the convective process. We
replace this dependency by a free dimensionless parameter, the mixing
length ($L$), assuming that bubbles rise a distance, $H_{\rm P},$ which is
proportional to  the atmospheric scale length of the halo at
each point:
\be
H_{\rm P}=L\frac{P}{\rho\,g}\, ,
\ee
with $g=GM/r^2$.
A typical acceleration (assuming isobaric perturbations) is:
\be
a=g\frac{\delta \rho}{\rho}=\frac{g}{\rho}\left(\frac{\partial
    \rho}{\partial S}\right)_{\rm P}\frac{\partial S}{\partial r}\,H_{\rm P}\, ,
\ee
so a typical bubble velocity is:
\be
v=\left(2\,a\,H_{\rm P}\right)^{1/2}=H_{\rm P}\left[2\,\frac{g}{\rho}\left(\frac{\partial
    \rho}{\partial S}\right)_{\rm P}\frac{\partial S}{\partial
  r}\right]^{1/2}\, .
\label{eq:v}
\ee
Once the velocity exceeds the local sound of speed in the halo, shocks
are created which quickly act to mix the bubble with its
surrounding. In the following calculation the velocity is not
allowed to exceed the speed of sound, $c_{\rm s}$, and in that case $H_{\rm P}$ is
reduced until $v=c_{\rm s}$ in \equ{v}. A {\bf maximal convection} model is
a model with arbitrary high $L$, so effectively the bubbles always
accelerate until the speed of sound, at which time they are broken and
mixed.

The Flux of energy per unit surface per unit mass is:
\be
F_c=C_{\rm P}\,v\left[\Delta \nabla T\,
  H_{\rm P}\right]=C_{\rm P}\,v\left[\frac{2}{5}\,\hat{\mu}\,T\frac{\partial S}{\partial
  r}\,H_{\rm P}\right]\, ,
\label{eq:flux}
\ee
which is determined by the halo profile from the simulation at each
time, and the mixing length $L$. $C_{\rm P}$
is the constant pressure heat capacity and is related to $\hat{\mu}$
by $C_{\rm P}=5/(2\hat{\mu}).$

A numerical solution of the mixing length model requires evaluation
of the incoming and outgoing fluxes from the boundaries of each radial
shell. The fluxes depend on the temperature gradient between each shell and the
ones directly below and above it, and interpolation of thermodynamic
properties from the shell centres to the shell's edges is required. A
solution using an explicit numerical scheme (with the fluxes
determined at the beginning of each timestep)  requires extremely small 
timesteps to avoid negative temperatures, so an implicit scheme which solves simultaneously for all
the temperatures and fluxes at the end of each timestep in each
convective area was implemented. This is done by inversion of the 
three-diagonal matrix which is obtained by discretization of \equ{flux}. 

In the context of the clump heating discussed in this paper, it should
be emphasized that the convection is not between the cold gas in the
clumps and the hot surrounding. It is only within the hot component,
and is the result of heating and cooling of that component. The
convection model assumes linear perturbations within a single gas phase which
separates into two phases (hot buoyant bubbles, and  cold sinking
gas). The propagation of clumps, which have a typical over-densities 
of $\gsim 1000$ is followed explicitly using the processes described
in \se{drag} - \se{destruction}. 

The ICM is mildly magnetized, with the non-thermal magnetic pressure
contributing at most $10\%$ of the total pressure \citep{churazov08}. 
This effect could lead to heat-flux buoyant instability \citep[HBI;
][]{parrish09} even
when the entropy profile is monotonically increasing provided there is
a temperature inversion near the core ($dT/dr>0$). These instabilities
act to align the magnetic 
fields perpendicular to the temperature gradient, in a manner that
suppresses further conduction \citep{parrish09}. Convective motions
are also somewhat suppressed even for material that
is hydrodynamically convectively unstable ($ds/dr<0$). 
Future work could, and should, use a revised
mixing length theory for which the driver of convection is temperature
inversion rather than entropy inversion. Such a model would need to
take into account the saturation of the instability as the magnetic
fields align themselves, baring in mind that heating by clumps is closely
related to turbulence driving by clumps. Also, the largely reduced convection
strength that is eluded to in \citep{parrish08b, parrish09}\footnote{See, for
example, the short-dashed lines in fig. 6 and 10 of \citet{parrish08b}.}
must be evaluated. Ultimately, the convection here is invoked to
smooth over local instabilities for which, at least according to the
spherical calculations, steep gradients of more than an order of
magnitude in temperature and entropy form at the spatial resolution
limit (thin red line of \fig{ent}). These extreme gradients (that are also present at edges of
radio bubbles in clusters) are far from linear perturbations, and the
validity of the linear analysis of the various convection
prescriptions is highly questionable. 

\subsection{The Observed Properties of Hot Gas with Cold Clumps}
\label{sec:effective}
In the current implementation, when clumps are destroyed, their mass
is added to the hot 
component instantly. In reality, the process of KH fragmentation,
followed by small scale evaporation and conduction of the debris will
yield a multiphased gas, with an effective entropy and temperature
which is between the values of the hot and cold phases. 
As will be shown later (\se{ncl}) , the clump density and clump
destruction rate increase 
towards the centre so effective cooler and lower entropy
values are expected there. The radiative signature of gas
heating through all the temperatures between $10^4K$ to the cluster
ambient gas temperature of $\sim 3\times 10^7K$ is expected to be
significantly different from that of radiative cooling since it is
governed by heating processes \citep[emission spectrum from heating
  gas, albeit by other heating mechanisms have been studies in][]{voit97,oh04}.
A framework of heating and cooling processes
in layers between hot and cold media have been proposed  by
\citet{begelman90,gnat10}. 
 
The observational signature of clump break up
would require detailed 3D 
simulation of clump interactions with cluster core gas,
and multiphased modeling of the radiative signature during the
heating process, and is beyond the scope of this work.
Instead,
we will plot below  the mass weighted entropy and temperature of the
two components. This is a lower limit for the observed entropy and
temperature as the  clumps' contribution to the luminosity,
particularly at X-ray wavelength, is probably small. Physically, it
corresponds to the thermodynamic properties expected in the event of
full mixing between the cold and hot phase. The actual temperature and
profile expected from the multiphase gas is thus bracketed between the
hot only component, and the mass weighting between the hot and cold
components presented in figs. \ref{fig:ent} and \ref{fig:temp}.

%%%%%%%%%%%%%%%%%%%%%%%%%%%%%%%%%
\section{Results}
\label{sec:results}

The simplified model described in \se{numerics} spans a multidimensional
parameter space including the
fraction of accreted gas in clumps, the initial clump masses, the number
of fragments that a clump breaks up to, and the mixing-length parameter for
convection.
In the absence of additional physical insight concerning
the formation mechanism and properties of these
clumps, we attempt to find a working set of values for the model parameters,
to serve as a feasibility test and hopefully provide clues for
acceptable values of the key parameters.
Ultimately, a more systematic survey of parameter space will have to be
conducted, with physically motivated values for key parameters such as
number and mass of clumps.

In this section we restrict ourselves to one typical CC cluster halo
with virial mass of $3\times 10^{14}\msun$ by $z=0$, a diffuse
baryon fraction of $10\%$,
and a smooth accretion history according to the average growth rate
of the main progenitor a la \citet{neistein06} \citep[see ][for a
detailed description]{bdn07}. The metallicity is assumed to be 
constant at $Z=0.3\,Z_\odot$, and the cooling is Bremsstrahlung
and line cooling according to \citet{sutherland93}. 
The initial resolution is $2000$ baryonic shells and $10,000$
dark-matter shells, roughly logarithmically spaced in their
initial radii. When shells expand, the adaptive mesh refinement algorithm 
splits them (\se{amr}), so the
resolution near the cluster core 
($\sim 50\kpc$) at all times is better than
$\sim 2\kpc$. This yields converged results in terms
of the profiles and the amount of gas that cools. 
We implement three different models for clump heating, as follows:

\begin{enumerate}
\item Model \mnc is the null model with no clump heating and no convection. 
It is meant to to reproduce the over-cooling problem.
\item Model \mcl adds clump heating but no convection, so it should show the 
over-heating instability. 
The fraction of baryons in clumps is $5\%$ and the clump initial mass is 
$10^8\msun$. 
The clumps are simulated by $10,000$ clump shells (\se{drag}). 
\item Model \mmc has the same clump heating as in \mcl but with maximum 
convection turned on (\se{mlt}). 
The smoothing by convection is supposed to regulate 
the clump heating and yield relaxed clusters compatible with observations.
\end{enumerate}  

\begin{figure*}
\includegraphics[width=7.5in, trim=50 0 0 0, clip=true]{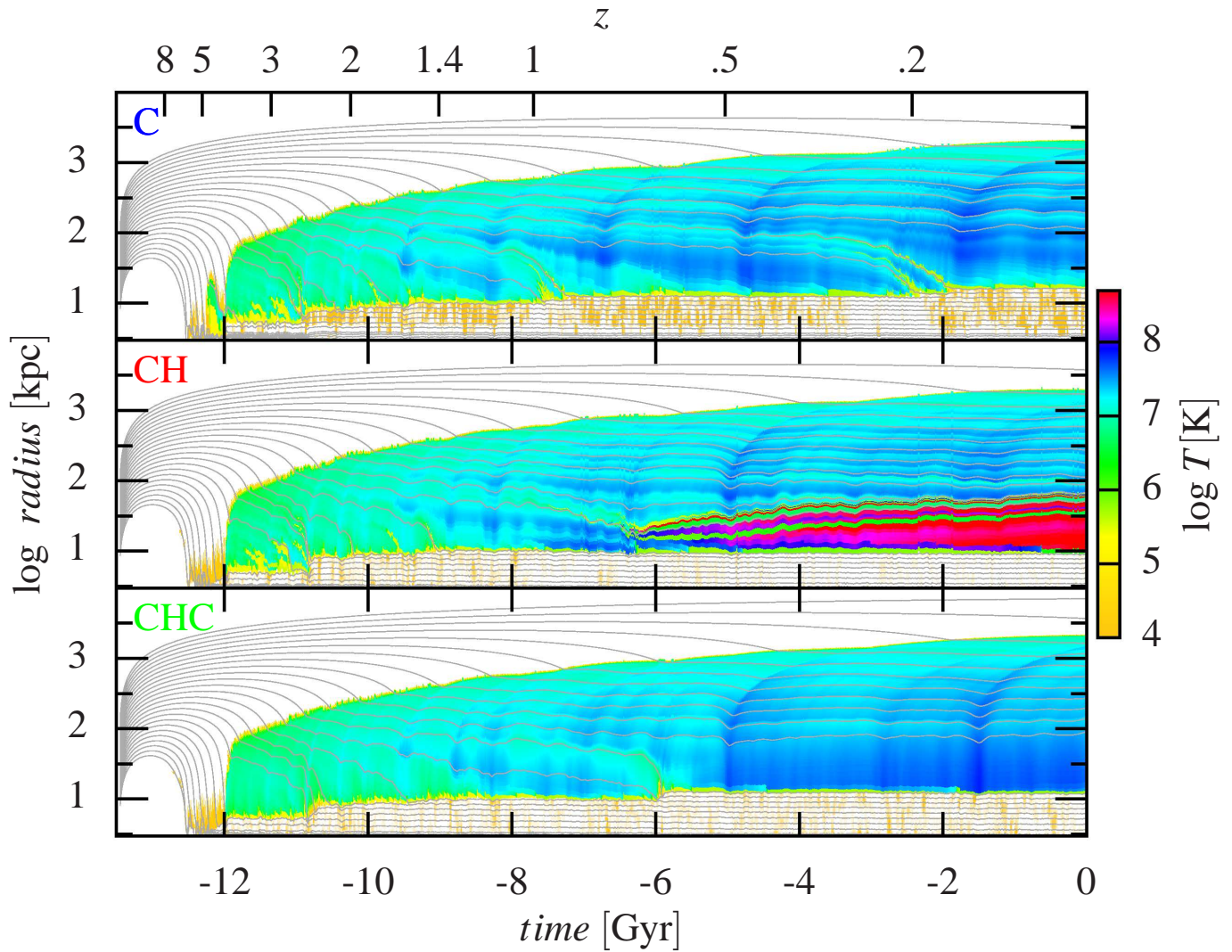}
\caption{\label{fig:3pan} 
The time evolution of the gas in a cluster for the three different models 
of clump heating. 
Shown in grey lines are the evolving radii of 25 spheres that encompass 
fixed baryon masses equally log-spaced in the range 
$10^{10} - 5\times 10^{13}\msun$.
The dark matter and clumps are not shown.
The final halo mass is $3\times 10^{14}\msun$.
Look-back time and redshift are marked at the bottom and at the top.
The color refers to gas temperature.
Top: Model C, cooling only.
Middle: Model CH, cooling and heating.
Bottom: Model CHC, cooling, heating and convection.
Model C shows cooling flows in the core between $10$ and $100\kpc$
during the last $6\Gyr$ (in particular near $t \sim -2\Gyr$). 
Model CH shows core over-heating and expansion in the last $6\Gyr$,
and Model CHC demonstrates how convection regulates the heating and brings
the cluster to an equilibrium.
}
\end{figure*}

\Fig{3pan} shows the time evolution of the gas in our simulated cluster
comparing the three different models for cooling, clump heating and convection.
The initial Hubble expansion and consequent turnaround of the Lagrangian gas 
shells is clearly seen, and the virial shock can be easily identified
after a collapse by a factor of $\sim 2$,
both by a jump in temperature from below $10^4$K to above $10^7$K,
and by the abrupt slow down of the infall velocity, which is almost brought to
a halt behind the shock. 
The global large-scale properties of the cluster are not affected by the
addition of heating and convection.
The virial radius evolves in a similar way, and
the typical temperature in the halo at $z=0$ remains at 
$T\sim 2-3\times 10^7$K ($\sim 2-3~\keV$), 
consistent with the expected virial temperature of $2.2\times 10^7$K 
for a cluster of virial mass $3\times 10^{14}\msun$.
However, the models differ at the core, within the innermost $100\kpc$
especially during the last $6\Gyr$ of evolution.
Model C shows inward cooling flows at all times, as expected \citep{fabian94}. 
With the addition of clump heating in Model CH
The cooling flows are stopped before $t=-6\Gyr$, 
the gas in certain shells is over-heated to extreme temperatures $\gsim 10^8$K,
these shells are interlaced with cooler shells of $\sim 10^6$K,
and together the whole core inflates. This behavior is in conflict with the
relaxed nature and smooth entropy and temperature profiles of CC 
clusters \citep{donahue06}.
The addition of convection in Model CHC removes the local over-heating,
and keeps the core in equilibrium at the virial temperature with no cooling
flow.
A more detailed comparison of the models and observations follows.
   
\subsection{Time Evolution of X-ray Luminosities and BCG Masses}
\begin{figure}
\begin{center}
\includegraphics[trim=10 0 10 0,clip=true,width=3.5in]{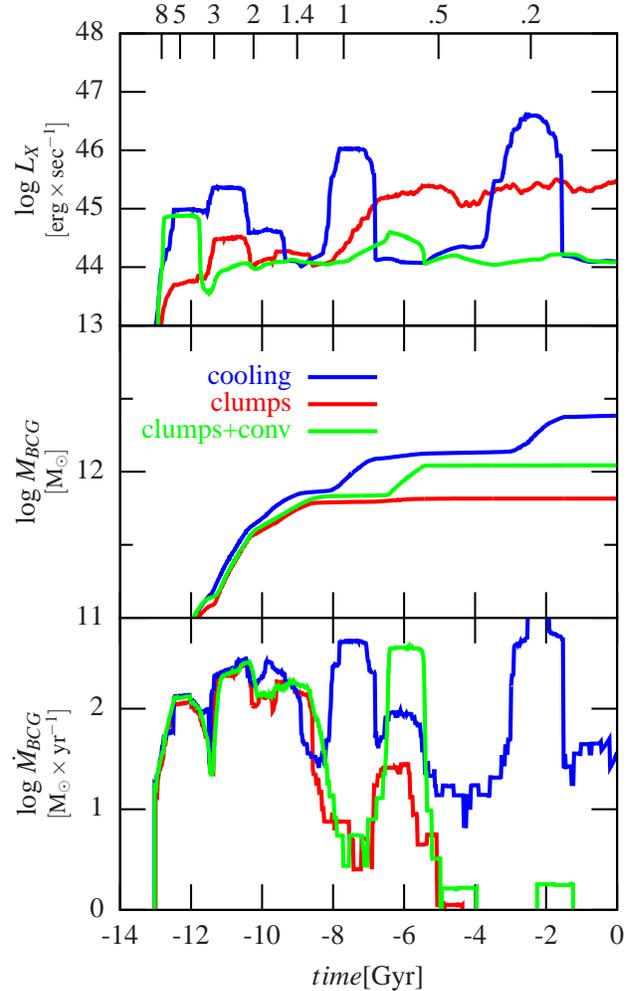}
\caption{\label{fig:time} 
Time evolution of X-ray luminosity (top), 
BCG mass (middle), and BCG accretion rate (bottom), 
all smoothed over $1\Gyr$. 
}
\end{center}
\end{figure}

\Fig{time} shows the evolution of X-ray luminosity,
the mass of the BCG, and the accretion rate onto it,
for the three different models.
The BCG is represented by the mass in the ``disc", the mass that is
 supported by angular momentum in the 1D simulations, extending to $\sim
10\kpc$.
The X-ray luminosity is obtained as the total cooling radiation from the
gas outside the BCG\footnote{Most of the energy is emitted from cooling 
of $\sim \keV$ gas, and is predicted to contribute to the X-ray 
luminosity that is relevant to the $L_x-T$ relation 
\citep[for example: ][]{markevitch98}.}.
The quantities are smoothed over $1\Gyr$ to erase sharp features 
that result from the discreteness of the calculation in the idealized
spherical calculation. 

Model C shows a variable luminosity, which occasionally exceeds 
by an order of magnitude or more the 
observed luminosity of $10^{44}-10^{45}\ergsec$
as derived from the $L_x-T$ relation
\citep{edge90,david93,markevitch98}.
In Model C the final BCG mass exceeds $2\times 10^{12}\msun$ 
and the accretion rate, which is an indicator for the star formation rate, 
has long episodes where it is in the range $100-1000\sy$,
at odds with observed CC clusters. 
When clump heating is added in Model CH, the cooling flow into the BCG
is drastically suppressed since before $t \sim -8\Gyr$, and it reaches a 
complete shutdown after $t \sim -5\Gyr$. 
The luminosity maintains a high level of $\sim 2\times 10^{45}\ergsec$ since $t \sim -7\Gyr$.
This results from the over-heating instability in the halo core,
leading to very dense shells that boost the dissipation due to drag 
interaction with the clumps and enhance the resulting radiation. 
The addition of convection in Model CHC brings the BCG mass to
a constant value of $\sim 10^{12}\msun$ with no detectable cooling flow
since $t\sim -6\Gyr$. The smoothing of the instability brings the luminosity to
a low value of $\sim 10^{44}\ergsec$, consistent with the
observed $L_X-T$ relation.

%--------------------
\subsection{Entropy and Temperature Profiles}

\begin{figure}
\includegraphics[trim=50 0 10 0,clip=true,width=3.5in]{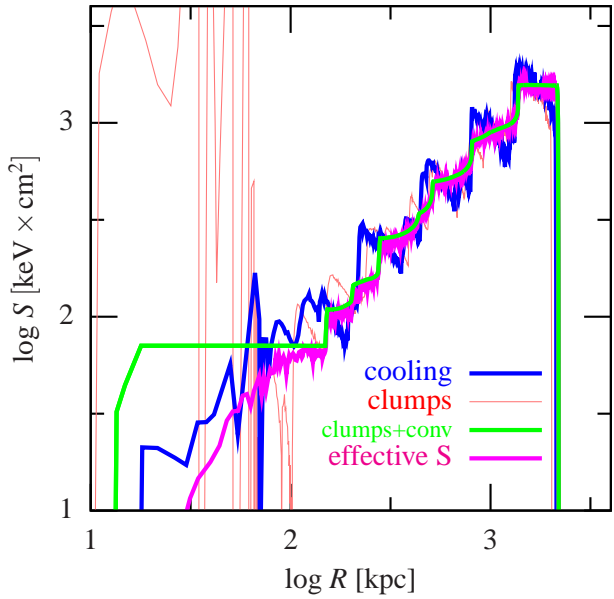}
\caption{\label{fig:ent} 
Radial profiles of entropy at $z=0$ for models C, CH and CHC.
Shown in three curves is the entropy of the hot medium, 
and an additional curve marked CHC$_{\rm eff}$ refers to the effective entropy
of the mixture of hot gas and the surviving cold clumps, computed as a
mass-weighted average and brackets the predicted entropy from below.
}
\end{figure}

\begin{figure}
\includegraphics[trim=10 0 10 0,clip=true,width=3.5in]{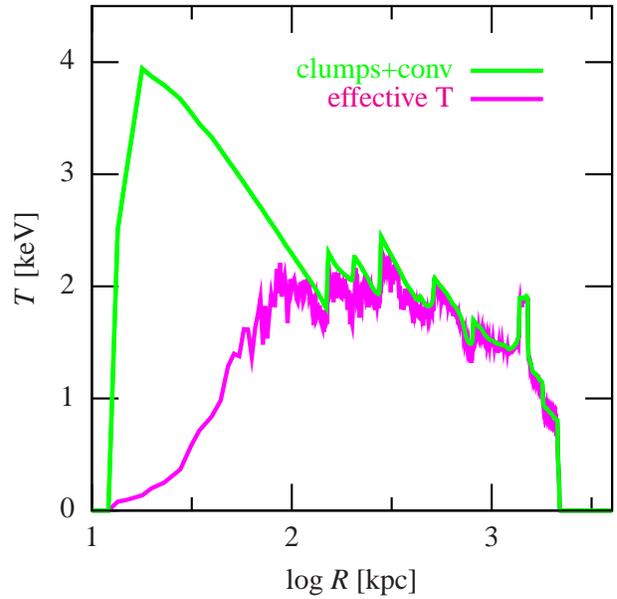}
\caption{\label{fig:temp} 
Radial profiles of temperature at $z=0$ for model CHC,
for the hot medium and for the mixture of hot medium and cold clumps,
computed as a mass-weighted average. The predicted temperature is
bracketed between this upper and lower limit.
}
\end{figure}

The $z=0$ entropy profiles of the three models are
plotted in \fig{ent} and the temperature profile of model \mmc is
plotted in \fig{temp}. 
The entropy profiles in the outer halo, overall increasing 
close to linearly 
with radius and having values $\sim 100\keV\times{ \rm cm}^2$ at
$100\kpc$, are similar in all models and consistent with observations
of CC clusters \citep{donahue06,cavagnolo09}. 
The apparent periodic fluctuations represent cold fronts that result from 
mergers of outward-propagating shocks at regular intervals with 
the virial shock (also visible in \fig{3pan}), which 
have been studied in \citet{birnboim10}. 
While the fluctuations in Models C and CH are locally 
non-monotonic, the entropy profile of Model CHC is monotonically increasing 
throughout because the convection removes negative entropy gradients.
Model CH shows a high-entropy core due to overheating,
and a strong variability representing a mixture of cold-dense and hot-dilute
shells that result from the overheating instability,
both in conflict with observations.
The convection introduced in Model CHC removes the fluctuations
and produces what seems to be a flat core entropy profile inside 
$150\kpc$ at $50\keV\times{\rm cm}^2$. Such a core is still 
inconsistent with CC cluster cores, but recall that the profile shown is
limited to the entropy of the hot component alone.
The effective entropy that could actually be observed 
is a mixture of the entropy in the hot gas, cold clumps, and anything
in between as discussed in \se{effective}. 
The effective entropy profile shown in \fig{ent} is monotonically
increasing in the core down to $\sim 30\kpc$, consistent with CC clusters.

The temperature profile of Model CHC at $z=0$ is plotted in
\fig{temp}. It shows a roughly isothermal halo at the virial 
temperature outside the core of $\sim 100\kpc$, with a mild decline 
toward the virial radius,
as observed \citep{donahue06}.
This large-scale temperature profile has not been affected much by 
clump heating and convection.
The temperature of the hot component is rising toward the centre, by a factor
of $\sim 2$, but the effective mass-weighted temperature of the mixture of 
hot and cold components is declining toward the centre, by a factor of a few.
This mass-weighted effective temperature (that corresponds to the
single temperature the gas would if the phases were fully mixed) is a
lower limit to the luminosity-weighted temperatures observed, and is consistent with the
moderate temperature decline in CC cluster cores. The temperature profiles of
the various models, and its time evolution, can also be seen in \fig{3pan}.
 
\subsection{Sensitivity to choice of parameters}
\begin{figure}
\begin{center}
\includegraphics[trim=10 0 10 0,clip=true,width=3.5in]{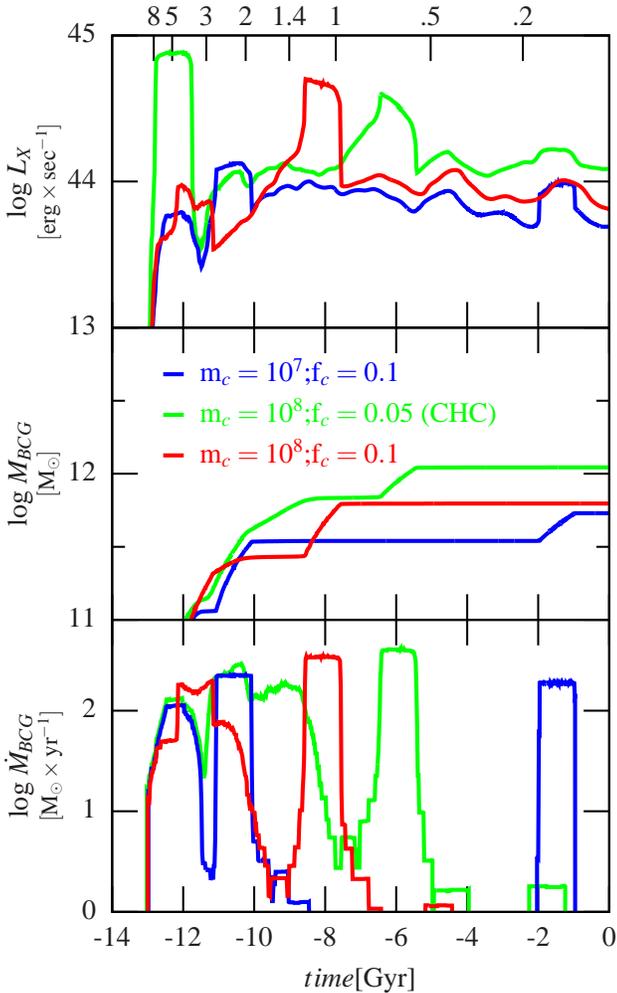}
\caption{
\label{fig:sensitivity} 
Time evolution of X-ray luminosity (top), 
BCG mass (middle), and BCG accretion rate (bottom), 
all smoothed over $1\Gyr$ of three sets of parameters. 
}
\end{center}
\end{figure}
The results presented in this section indicate that clumps of
$m_{c}=10^8\msun$ that make for $f_c=0.05$ of the gaseous component of clusters is
sufficient to remedy the over-cooling problem, and quench the
accretion of gas onto the BCG. These parameters were picked to
demonstrate the effectiveness of our model. We find that our simulated
clusters are not particularly 
sensitive to these values, and that no fine tuning is
required. Rather, a wide envelope of allowed parameters is allowed. \fig{sensitivity} is analogue to \fig{time} and compares
the luminosity, BCG mass and accretion rate predicted by simulations
with maximal convection for a parameter choice of:
$(m_{c}=10^7\msun,f_{c}=0.1)$ and $(m_{c}=10^8\msun,f_{c}=0.1)$, along
with our fiducial model of
$(m_{c}=10^8\msun,f_{c}=0.05)$. The same
behaviour is also found in the resulting radial profiles. A simulation
with the parameters  $(m_{c}=10^7\msun,f_{c}=0.05)$ allowed for too
much cooling, and final BCG mass of $1.5\times 10^{12}\msun$ - 
that is slightly excessive. 

Beside these parameters, some of the theoretical model assumptions
(for example the drag efficiency, the fragmentation of clumps and the
convection) might need to be modified after more detailed 3D simulations or additional
observational  constraints are found. We expect that for a different model the
parameters will need to be readjusted, but since the heating model is
not particularly sensitive, we expect that such a choice will always
be possible. We do
not believe that a comprehensive parameter survey will be beneficial
at this point, until the different components of this model are better
constraint either theoretically or observationally.  

%%%%%%%%%%%%%%%%%%%%%%%%%%%%%%
\section{Observational Signatures of Clump Heating}
\label{sec:ncl}

The model parameters chosen in this paper were motivated by the
analysis of DB08 and were calibrated such that the model crudely reproduces
the properties of CC clusters in order to demonstrate the feasibility of such a
model of gravitational heating. The reproduced properties include
the BCG mass, the cold mass accretion history, the X-ray luminosity and
the entropy and temperature profiles.
This model makes additional predictions that could distinguish it from 
other heating models such as the ones based on AGN feedback. 
Some of these predictions are discussed here.

%-----------------------------
\subsection{Cold gas in the ICM}
\begin{figure}
\includegraphics[width=3.5,trim=30  0 20 0,clip,width=3.5in]{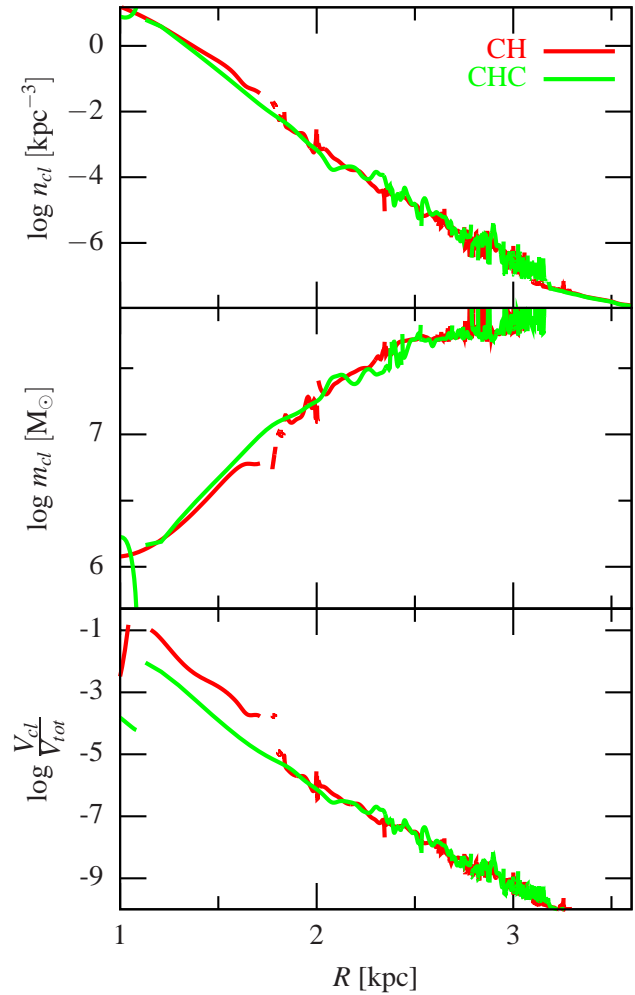}
\caption{\label{fig:ncl} 
Average number density of clumps (top), average clump mass (middle) and
fractional volume in clumps (bottom)
as a function of radius at $z=0$ for Models CH and CHC.
}
\end{figure}
The mass function of clumps at every radius is a distinctive
prediction of our proposed model.
Assuming an initially uniform population of $10^8\msun$ clumps,
\fig{ncl} shows the profiles of number density and average
clump mass for Models CH and CHC.
The fraction of volume occupied by cold clumps is shown in the bottom panel.
These predictions depend on the specific choice of the initial
mass of the clumps and on the baryonic fraction of mass in these
clumps. The number density of clumps increases toward the centre
$\prop r^{-2}$, partly reflecting the general density profile of the cluster 
and partly because of the break up of clumps into fragments 
(\se{clump_numerics}). 
Once clumps are fully destroyed 
(fragments $< 10^4\msun$), their mass is added
to the hot component, and they are no longer plotted in \fig{ncl}. The
figure thus shows the steady state population of clumps as they are
continuously accreted and destroyed.
The average clump mass is declining from the initial value of $10^8\msun$ 
in the outer halo to $\sim 10^6\msun$ near the centre, reflecting the
clump fragmentation as they flow in.
The fraction of volume occupied by the cold clumps peaks at a few percent 
near the cluster core, and drops to smaller values at larger radii. 
This justifies ignoring the additional pressure caused by this component.

The clumps are initially cold, and, except for the mild compression they 
undergo as they fall in following the increasing pressure of the ambient gas, 
they are not expected to heat up or emit much radiation.
However, as the clumps are disrupted by hydrodynamical instabilities
and possibly also by tidal effects, they
fragment into smaller pieces for which conduction and 
evaporation become more important (\se{effective}).
Once heated to intermediate temperatures, the gas begins to radiate.  
Spectroscopic observations could in principle constrain the validity of 
this model in comparison with AGN-feedback models, where one expects
gas cooling rather than heating through the intermediate temperatures.
The current simplified implementation of clump heating does not permit a 
proper comparison, which is left for future work. 
Additionally, three-dimensional simulations are required for a detailed 
analysis of the shape of the clumps 
as they are stretched 
perhaps leading to morphologies resembling filaments \citep{murray04}. 
This emission, in $H_\alpha$
and line and continuum emission of the intermittent X-ray temperature gas,
may allow more accurate comparisons of this model with the 
observed profiles in cluster cores. 
Observations in the Perseus cluster (NGC 1275)
\citep{conselice01,fabian08} show a complicated
structure of $H_\alpha$ filaments and blobs. The typical masses of these 
features are $10^6-10^8\msun$, consistent with the allowed mass range
for clumps in DB08, and with the
distribution predicted by our model (\fig{ncl}). We note that this
result depends on the initial mass of the clumps - a free parameter
here. The consistency of this prediction with observation is an
indication that our choice of initial mass of $10^8\msun$ is reasonable.
\citet{fabian08} invoked strong magnetic fields to stabilize the filaments for
cosmological times, such that their age can match that of the observed
radio bubbles. 
Our heating model suggests instead that these filaments are constantly 
being destroyed, as new clumps enter the cluster core, 
get stretched and destroyed, and create new filaments. 
The projected filling factor of
these structures approaches unity within the innermost $10\kpc$
and it drops outwards \citep{conselice01}. Such a behaviour is predicted
by our model (\fig{ncl}). $H_\alpha$ emission in other clusters have been
reported by \citet{heckman89}, who found that the cold gas has
velocities at random directions rather than a coherent radial cooling-flow
pattern. This kinematics could be interpreted as clumps oscillating 
in and out at the vicinity of the BCG. 
Structures of neutral gas are also seen in the Virgo Cluster by the
ALFALFA 21cm survey \citep{giovanelli07,kent07} showing evidence for
neutral gas arranged in clumps, with masses as low as the
detection limit of $2\times 10^7\msun$, sometimes with no optical counterparts. 

%-----------------------
\subsection{Turbulence in the ICM}

\begin{figure}
\includegraphics[width=3.5,trim=0  0 20 0,clip,width=3.5in]{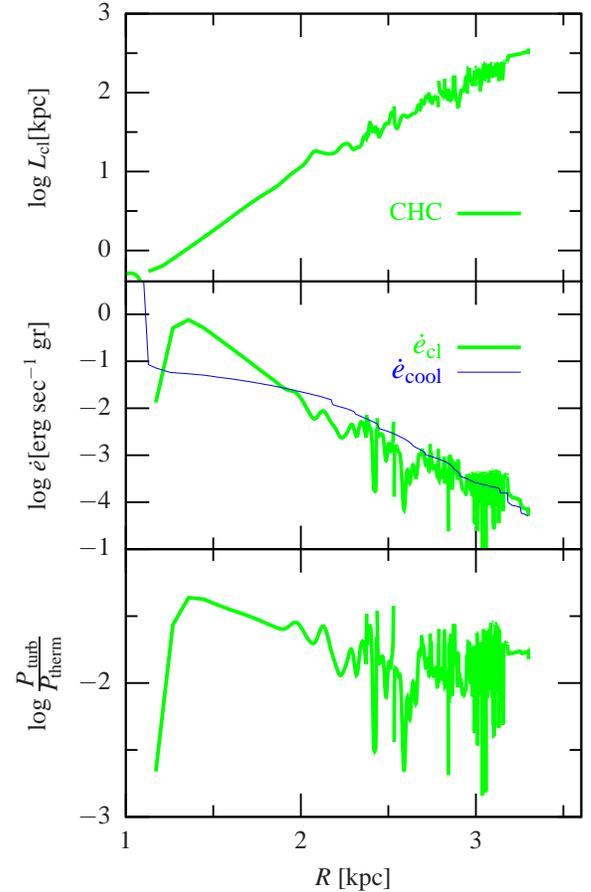}
\caption{\label{fig:turb}
The average distance between clumps  (top), the average drag energy
injection compared to the cooling rate (middle) and the ratio of turbulent pressure to thermal
pressure (bottom) all  as a function of radius
at $z=0$ for Model CHC.
}
\end{figure}

Another prediction of this model is the power of turbulence
that is produced in the ICM.
When the clump velocities are subsonic with respect to the ICM, the
energy and momentum of the drag are 
first converted to kinetic energy in turbulence, which cascades down to
smaller scalelengths where it dissipates into heat. When the motion
is supersonic, some of the energy is converted directly into heat, but
since momentum is conserved, some of the energy must be transferred as
kinetic energy to the ICM. 
We assume that the turbulence is generated on a scalelength comparable to the
average distance between clumps, $L$, and that a Kolmogorov spectrum 
governs the cascade of eddies from this scale 
to smaller scales. The turbulent energy and pressure 
are then 
\begin{align}
e_{\rm turb}&=\frac{3}{2}\,c_0(\dot{e}_{\rm drag}\,L)^{2/3} \, ,\\
P_{\rm turb}&=(\gamma-1)\,\rho_{\rm ICM}\,e_{\rm turb}=c_0\,\rho_{\rm ICM}(\dot{e}_{\rm drag}\,L)^{2/3}\, ,\nonumber
\end{align}
\citep{landau59}, with $c_0=2.1$ taken from \citet{popo00}. 
The amplitude of the turbulence spectrum, and the turbulent pressure can thus be determined
once we know the average distance between clumps and the drag heating rate
at every radius. 
The ratio between the turbulent pressure and the thermal pressure for
Model CHC is plotted in \fig{turb}. 
The level of turbulent pressure in the cluster core is at the level of 
$\sim 6\%$ of the thermal pressure,
in qualitative agreement with observations
\citep{rebusco06,churazov08}. The average clump distance, $L$ and the
energy injection rate $\dot{e}_{\rm drag}$ are separately shown is
\fig{turb} as well. The value of $L$ at the center is in rough
agreement with the $H_\alpha$ structures observed in
\citet{conselice01} (further discussed below) and the drag heating
exceeds the cooling rate near the centre, but becomes comparable
around the $100\kpc$, consistent with convection occurring near the
centre as suggested in \se{mlt_numerics}.
%------------------------
\subsection{High Velocity Clouds in the Galactic Halo}

High velocity clouds (HVCs) are observed in 21cm HI data
\citep{blitz99} as concentrations of gas moving at
velocities $>100{\rm km\,s^{-1}}$ relative to the rotating frame of the
Milky Way.  Options for the spatial origin of the HVCs
range from the Milky Way (MW) disc \citep[][and references therein]{wakker08}, 
through the Magellanic clouds \citep{olano08}, 
to extragalactic origin \citep{blitz99}. 
The distance to and the ionization fraction of individual clouds 
(and therefore their size and mass) are unknown. 
\citet{putman03} estimated distances of HVCs based
on their H$_\alpha$ flux and models for the emission of ionizing
radiation from the MW. They find, within the modeling and measurement
uncertainties, that most HVCs are within a distance of $\lesssim
30\kpc$, indicating a mass range of $10^4-10^8\msun,$
\citep{putman03,birnboim09}. Other estimates by absorption
features \citep{thom08} yield comparable distances and masses. 
The origin of the HVCs is unclear. Models have
proposed that they form within the Galactic halo by cooling
instabilities \citep{maller04,keres09}. See, however, \citet{fraternali08}. 
\citet{binney09} analyzed formation of clumps in smooth MW halo conditions from 
thermal instability and concluded 
that it is unlikely to occur near the centre. They find, however, that
the conditions become favorable closer to the halo virial radius
and when the entropy profile is shallow \citep[as is shown numerically
in ][]{kaufmann09}. Their stability analysis
tests growth of instability from infinitesimal perturbations but the
growth of non-linear perturbations caused by shocks, collisions,
and gravitational perturbers depends on initial conditions.
The line-emission peak of the cooling curve at $\sim 10^5K$ 
would make the warm cosmic filaments outside clusters
more susceptible for clump formation.
However, clumps have been shown to be a natural consequence of cold-flow 
filament breakup by hydrodynamic instabilities \citep{keres09}.
For example, streams that do not flow radially to the halo centre
are susceptible to Rayleigh-Taylor instability. 
Also, shocks that originate from the galaxy, e.g., by mergers or by starbusts,
are likely to form clumps by Richtmyer-Meshkov instability.
The destruction of these clumps, and their interaction with the Galaxy and
the IGM, are likewise under debate \citep{fraternali08,keres09}.

We point at an obvious analogy between the observed Galactic HVCs and the 
clumps addressed in this paper. Their masses are comparable, and their
spatial distributions in the halo and toward its centre are possibly similar.
The larger pressure in the ICM of a more massive halo
would make the cluster clumps denser than the Galactic HVCs 
(by the ratio of virial temperatures which is more than $10$ times
larger for clusters), so clumps of a similar mass could survive longer in
clusters, allowing them to travel to the centre according to
the estimates in \se{numerics}.
The total mass encompassed in HVCs seems to be $\gtrsim 10^9\msun$, 
making it a few percent of the total baryons in the MW halo,
in good agreement with our fiducial choice of parameters.
A missing piece of the model is the yet unspecified origin of the clumps,
in terms of physical mechanism and location.
The existence of HVCs provides circumstantial evidence that such clumps 
might form. As long as the clumps are formed before they fall into the halo,
or even if they form inside the halo at a radius that is not much smaller than
the virial radius \citep{maller04,keres09}, the gravitational energy that is
released during their infall is significantly larger than the
energy required for heating the clumps to the virial temperature (DB08).

%%%%%%%%%%%%%%%%
\section{Discussion and Conclusion}
\label{sec:discussion}

%gravitational heating
The concept of gravitational heating of ICM gas as a partial or full
solution to the cooling-flow problem is more general than the specific 
clump model discussed in this paper. It is easy to show that
the gravitational energy that is
released as baryonic matter falls in through the halo potential well
is enough to balance the cooling rates in groups within haloes of 
virial masses $\sim 10^{13}\msun$, and it exceeds the cooling rate 
by more than an order of magnitude in cluster haloes $\sim 10^{14-15}\msun$ 
(DB08).
This point has also been made in \citet{fabian03,wang08,khochfar08}.
\citet{elzant04b,faltenbacher07b} and \citet{khochfar08}
tap into the same energy source, but utilize dynamical friction that is
less effective. \citet{naab07,johansson09} show,
however, that dynamical friction can efficiently stop gas accretion
onto massive elliptical galaxies.
Conduction also taps into this energy source, and was proposed  three
decades ago 
\citep{bertschinger86b,rosner89}. It is probably ruled out
because of magnetic field suppression of the conduction
\citep[][ and reference within]{binney81, fabian94}, and because the
resulting profiles would have a flat temperature core
\citep{bregman88}. See, however,  \citet{narayan01} and \citet{kim03} for alternative ideas.
Here we use the same energy source, but the physical process used to
couple the energy with the baryonic cooling component is hydrodynamic
drag for which the strength of interaction peaks at low clump masses and
high velocities. 

The challenge for every heating model is to distribute the energy
uniformly throughout the cluster core both in space and time 
\citep{deyoung08,cattaneo09}. 
In order to obey the observational constraints, 
the heating mechanism should suppress the gas mass that actually cools
by two orders of magnitude.  
Such a shutdown requires that the mechanism should act smoothly
over a scale of a few kpc, set by the
smallest object that would cool in the absence of feedback while
being continuously heated by conduction from its surrounding. 
This scale follows from
\be
L_{\rm cond}\sim \sqrt{\eta\,\kappa_{\rm
    Sp}\,t}=7\sqrt{\eta_{0.2}\,T_2^{5/2}\,n_{-2}^{-1}\,t_8}\kpc
\ee
with $\kappa_{\rm Sp}$ the Spitzer coefficient \citep{spitzer62},
$\eta$ the reduction of the Spitzer coefficient due to magnetic fields,
$t$ the cooling time of the gas, and $T_2=kT/2\keV,~n_{-2}=n/10^{-2}\,{\rm
  cm}^{-3},~t_8=t/10^8\yr,~\eta_{0.2}=\eta/0.2\, .$ 
The value $\eta=0.2$ is a reasonable
upper limit for the efficiency of conduction
\citep{narayan01}.
It also has to be temporally smooth
over the cooling timescale, which is a few $10^8\yr$ at most
\citep{donahue06}. 
The fact that the energy source of gravitational infall is automatically 
distributed over the cluster volume and over Hubble times makes it easier 
to meet this challenge with gravitational heating than it is with
AGN-feedback models, where the source is on scales ten orders of magnitude 
smaller than the cluster scales. However, the coupling of the infalling baryons
with the ICM should be such that most of the gravitational energy 
is deposited in the cluster core.   

%clump heating summary
This paper addressed one specific scenario of gravitational heating, 
in which the mechanism for feeding the energy into the ambient ICM is via 
hydrodynamic drag acting on small clumps of cold gas.
In a typical cluster, with a modest gas fraction $\sim 5\%$ of the
accreted baryons  
in cold clumps of $\sim 10^8\msun$, the clump heating suppresses 
the cooling flows toward the BCG for the last $6\Gyr$.  The
conditions at the core do not affect the incoming clumps,
so the process is not strictly self-regulating. Regardless, we find that
the large-scale properties of the cluster, 
such as the overall gas fraction, virial shock and temperature outside 
the core, 
are unaffected by the heating.
Furthermore, the core does not explode; it  
reacts to the heating smoothly and quiescently 
without a need for inherent self regulation. 
Due to the over-heating, the central density decreases and the total
X-ray luminosity emitted from the core (\fig{time}) declines
such that the core obeys the observed $L_x-T$ relation. On cluster
core scales, convection acts to flatten the 
entropy profile of the hot component and carry heat outwards as expected. The
effective entropy profile  after taking into account the cold
clumps as well does not exhibit an entropy core, and is consistent
with observed entropy profiles, to within the model limitations that
are discussed. A local
instability caused by the linear dependence of the heating rate 
on density ($\dot{e}_{\rm heat}\sim \rho$) acts to create extreme entropy and
temperature peaks of sub-kpc scales. These peaks create strong
convection that, once 
accounted for in the simulations, stabilize the heating process on local
scales as well. 

With the fiducial values of the parameters used in this work in
a $3\times 10^{14}\msun$ cluster halo, the model is successful
in quenching the cooling flows and in reproducing adequate
BCGs, X-ray luminosities, and entropy and temperature profiles. 
The model also predicts the expected level of turbulence in clusters,
and the fraction of cold gas as a function of radius.
These two observables are predictions of this model,
while they are not naturally addressed by AGN-feedback models.

%dynamic effects
In DB08 we analyzed clump heating in a
static halo using a Monte-Carlo approach to simulate an ensemble of clump
trajectories. 
We realized that the heating rate in the core is higher than the cooling
rate, which could cause the core to expand.
In order to see how the cluster could reach a steady-state configuration
one must allow the system to respond dynamically.
The current implementation of the model using a 1D hydro code to simulate
a cluster in the cosmological context allows us to 
do just that. 
We find that convection can regulate the over-heating instability and
produce a cluster with no cooling flow in steady state.
The net effect of the dynamical response is to make the heating 
more efficient. For example, with 5\% of the baryons in $10^8\msun$ clumps
inflowing into a static $3\times 10^{14}\msun$ cluster, 
our estimates in DB08 indicates a heating to cooling ratio slightly
below unity, while here we find it to be above unity, 
predominantly due to the net expansion of the core.
The dynamical evolution of the cluster (\fig{3pan}) 
is noticeable especially from $z\sim 2$ and on, as
the core takes a different thermodynamic trajectory in response to the
heating.  The BCG mass is smaller, and the core density 
is lower than in the simulation without heating. It
is therefore possible that other heating mechanisms 
that were tested within a static framework 
\citep{conroy08,fabian03,kim03,kim05,ciotti07} would also show
different evolution tracks once the gas 
is allowed to dynamically adjust to the energy input while the halo is growing.

%origin, observability of clumps
The main missing piece in the proposed model of clumps as the agents for
depositing the gravitational energy of infall in the ICM core 
is the unspecified mechanism and birthplace for the formation of clumps
with the desired properties of abundance and mass.
While clumps probably do not form in-situ at the cores of clusters
\citep{binney09}, they can form within cosmic filaments 
\citep{dekel09,ceverino10,fumagalli11}
and in the edges
of haloes. Non-linear perturbations and complicated halo geometries
can stimulate the formation of such clumps \citep{keres09}. 
The analogy between the observed HVCs and the desired clumps for
heating clusters is promising and may provide a clue for the origin of
these clumps.

The degree of clumpiness needed for effective clump heating, at the level of 
$\sim 5\%$, does not seem to be very demanding. 
The clumps may be hard to detect outside the halo virial radius
as their temperatures are expected to be only slightly lower than the 
temperature of the surrounding filaments, and therefore their inner densities
in pressure equilibrium are expected to be only slightly higher.
The clumps become denser and possibly more detectable
once they enter the hotter and denser ICM, and especially as they
approach the cluster core.
We are encouraged by observations of $H_\alpha$
structures with masses of $10^6-10^8\msun$ around the Perseus BCG
\citep{fabian08} but a more careful comparison needs to be made,
addressing the ionization states and the radiative signature of clumps
as they are disrupted and heated \citep{begelman90,gnat10}. 
One should also work out the spectrum of the X-ray emission from
the multi-phased gas as it is heated by conduction and radiation. Here
we only provide upper and lower limits to the observed temperature and
entropy, derived by assuming either that only the hot component is observed,
and that full mixing occurs instantaneously as the clumps disintegrate.

%1D-3D
The exact details of clump-ICM interactions, and the response of the
ICM, cannot be properly addressed in 1D simulations where 
hydrodynamic instabilities are almost completely suppressed
independent of the quenching mechanism. 
In this first crude study we rely on the fact that
the model proposed here naturally deposits the
energy over $\sim 1\kpc$ scales (\fig{ncl}) in a continuous manner. 
Nevertheless, a realistic study of whether the gas 
cooling is sufficiently suppressed would require a proper 3D simulation
with clump heating implemented.

%Caveats of convection
The inherent runaway expansion that occurs when
heating is faster than cooling is 
damped by a 1D convection model, which introduces a free
mixing-length parameter that can only be calibrated by 3D
simulations. Our assumption here, that the convection is maximal in
the sense that bubbles accelerate until they reach the speed of sound may be
overly optimistic. Additionally, weak magnetic fields
in the ICM may affect the nature and strength of the convection and could alter
the general behaviour of convection to follow temperature inversions rather than
entropy inversions. We find that the results are not particularly sensitive
to the value of the mixing length parameter as the local perturbations
are short scaled and (at least within the framework of the model)
entropy inversion is erased for a wide range of mixing length
parameters. These effects must be addressed in
future work. 

In \se{3d} we outlined a proposed implementation of a 3D subgrid 
model for these clumps, which will allow all these issues to be
addressed. A different kind of 2D and 3D simulations, of
interactions between single clumps with the ICM gas, have been conducted in
the past \citep{murray04}, but not for the specific conditions in
clusters, and without some of the crucial physical components such as cooling
and conduction. 

Our results support the notion that gravitational heating by the 
instreaming baryons
could be a major player in the heating of the cores of massive galaxies
and clusters. Whether this mechanism by itself is sufficient for preventing 
cooling flows or it must work in concert with AGN feedback is yet
to be investigated using 3D cluster simulations
(e.g. Zinger et al., in preparation).

\section*{Acknowledgments}

We thank Ami Glasner for his advice on mixing-length theory. 
We acknowledge fruitful discussions with Peng Oh, Orly Gnat, Jerry
Ostriker, and Ian Parrish, and thank the referee for helpful comments.
This research has been partly supported by ISF grant 6/08,
by GIF grant G-1052-104.7/2009, by a DIP grant, and by NSF grant AST-1010033.
YB was an ITC Fellow at CfA, Harvard.

%%%%%%%%%%%%%%%%%%%%%%%%%%%%%%%%%%%%%%%%%%%%%
\bibliographystyle{mn2e}
\bibliography{yuval}
%%%%%%%%%%%%%%%%%%%%%%%%%%%%%%%%%%%%%%%%%%%%%

\label{lastpage}
\end{document}